\documentclass[aps,pra,superscriptaddress,preprint,12pt]{revtex4-2}

\usepackage{amsmath,amssymb}
\usepackage{bm}
\usepackage[dvipsnames]{xcolor}
\usepackage{graphicx}
\usepackage{CJKutf8}
\usepackage[normalem]{ulem}

\newcommand{\Qwp}{\mathcal{Q}_\mathrm{W}^p}

\def\mr{\mathrm}

\def\mc{\mathcal}

\begin{document}
\begin{CJK*}{UTF8}{gbsn}
\title{Feasibility of extracting the proton weak charge from
quantum-control measurements of atomic parity violation on the $2s-3s$ or $2s-4s$ transition in hydrogen}

\author{Jiguang Li (李冀光)}\email{li\_jiguang@iapcm.ac.cn}
\affiliation{Institute of Applied Physics and Computational Mathematics, Beijing 100088, China}

\author{Andrei Derevianko}\email{andrei@unr.edu}
\affiliation{Department of Physics, University of Nevada, Reno, Nevada 89557, USA}

\author{D. S. Elliott}\email{daniel.s.elliott.1@purdue.edu}
\affiliation{Department of Physics and Astronomy, Purdue University, West Lafayette, Indiana 47907, USA}
\affiliation{School of Electrical and Computer Engineering, Purdue University, West Lafayette, Indiana 47907, USA}
\affiliation{Purdue Quantum Science and Engineering Institute, Purdue University, West Lafayette, Indiana 47907, USA}

\date{\today}

\begin{abstract}
    We explore the feasibility of extracting electroweak observables from a measurement of atomic parity violation in hydrogen. Our proposed quantum-control scheme focuses on the $2s-3s$ or $2s-4s$ transitions in hydrogen. This work is motivated by the recently observed anomaly in the W-boson mass, which may substantially modify the Standard Model value of the proton weak charge. We also study the accuracy of the previously employed approximations in computing parity-violating effects in hydrogen.
\end{abstract}
\maketitle
\end{CJK*}

\section{Introduction}
\label{Sec:Intro}
Atomic parity violation (APV) is a powerful probe of the low-energy electroweak sector of the Standard Model (SM) of elementary particles. The APV results are both unique and  complementary to those from particle colliders, in particular because APV probes Z-boson exchange at low momentum transfer $Q$ ($Q \sim \hbar/r_p \approx 0.2 \, \mathrm{GeV}/c$ for a nucleus of  size $\sim r_p \sim \mathrm{fm}$).
The already demonstrated precision of table-top APV experiments and theoretical interpretations places important constraints on exotic beyond-SM physics; these are often competitive to those derived from colliders. The rich history of APV and its implications have been reviewed recently in Refs.~\cite{SafBudDeM2018.RMP,WiemanDerevianko-SM50-2019-arXiv}. We will use APV and parity non-conservation (PNC) interchangeably.

Microscopically, APV is caused by the weak interaction mediated by the exchange of a Z boson. 
The relevant contribution to the SM Hamiltonian density reads \cite{Mar95}
\begin{equation}
\mc{H}_\mathrm{PV} = \frac{{{G_\mathrm{F}}}}{{\sqrt 2 }}\sum_{q} {\left( {{C^{(1)}_{q}}\,\bar e{\gamma _\mu }{\gamma _5}e\,\bar q{\gamma ^\mu }q + {C^{(2)}_{q}}\,\bar e{\gamma _\mu }e\,\bar q{\gamma ^\mu }{\gamma _5}q} 
+ {C^{(3)}_{q}} \partial_\nu (\bar{e} \sigma^{\mu \nu} e) \bar{q} \gamma^\mu \gamma _5 q
\right)} \,, \label{Eq:HPV_density}
\end{equation}
where the Fermi constant $G_\mathrm{F} \approx  2.22254 \times10^{-14}  \,\mathrm{a.u.}$,
the summation is over quarks,  $e$ and  $q$ are field operators for electrons and quarks respectively, and $\gamma$'s are the conventional Dirac matrices~\cite{BjoDre64}.

The  coupling of the electron axial-vector currents to the quark vector currents is parameterized
by the constants $C_{q}^{(1)}$;
the constants
${C_{q}^{(2)}}$ describe the coupling of the electron vector currents to quark axial-vector currents.  The $C_{q}^{(1)}$ contributions  lead to nuclear-spin-independent interactions and $C_{q}^{(2)}$ --- to the nuclear-spin-dependent interactions.  The $C_{q}^{(3)}$ contribution is due to the ``anomalous weak moment'' and it also leads to nuclear-spin-dependent effects.

The weak Hamiltonian~\eqref{Eq:HPV_density} can be simplified further by lumping quarks into nucleons (e.g., a proton is composed of two up and one down quarks),
\begin{equation}
\mc{H}_\mathrm{PV,p} = \frac{{{G_\mathrm{F}}}}{{\sqrt 2 }}{\left( {{C^{(1)}_{p}}\,\bar e{\gamma _\mu }{\gamma _5}e\,\bar p{\gamma ^\mu }p + {C^{(2)}_{p}}\,\bar e{\gamma _\mu }e\,\bar p{\gamma ^\mu }{\gamma _5}p} 
+ {C^{(3)}_{p}} \partial_\nu (\bar{e} \sigma^{\mu \nu} e) \bar{p} \gamma^\mu \gamma _5 p
\right)} \,, \label{Eq:HPV_densityProton}
\end{equation}
where we omitted couplings to neutrons because of our interest in the hydrogen atom. In general, the couplings to protons and neutrons read~\cite{MarcianoSanda1978} 
\begin{subequations}
\begin{align}
C_{p}^{(1)}&=2C_{u}^{(1)}+C_{d}^{(1)} = \frac{1}{2}\left( {1 - 4\,{\sin }^2}{\theta_\mathrm{W}}\right), \label{Eq:C1p}\\ 
C_{n}^{(1)}&=C_{u}^{(1)}+2C_{d}^{(1)}=  - \frac{1}{2}, \\
C_p^{(2)} &=  - C_n^{(2)} = {g_A}C_p^{(1)},
\end{align}
\end{subequations}
reflecting the quark composition of nucleons. Here 
$\theta_\textrm{W}$ is the  Weinberg angle and $g_A \approx 1.28$ is the scale factor accounting for the partially conserved axial vector current. Finally, the anomalous weak moment contribution is suppressed by the small value of the fine-structure constant $\alpha$ as $C_p^{(3)} \approx \alpha/(2\pi) C_p^{(2)}$. Notice that the more conventional parameterization of the proton spin independent coupling $C_{p}^{(1)}$ is in terms of the proton weak charge $\Qwp$. At the tree level (i.e. excluding radiative corrections), 
 \begin{equation}
     \Qwp = 2C_{p}^{(1)} \,. \label{Eq:Cp1-Qwp}
 \end{equation}
 
Since $\sin^2\theta _\mathrm{W}  \approx 1/4$, the neutron contribution $C_n^{(1)}$  dominates $\mc{H}_\mathrm{PV}$ in all atoms except for the hydrogen atom. 
As a consequence, hydrogen is uniquely sensitive to $\sin^2\theta_\mr{W}$ potentially enabling its extraction  at the low momentum transfer~\cite{Erler2003,Androic2018}.
This is the motivation for our focus on the hydrogen atom, as the experiments with heavy atoms, such as cesium~\cite{WooBenCho97} or ytterbium~\cite{Antypas2018.APV-Yb},
are predominantly sensitive to the neutron coupling constant $C_{n}^{(1)}$.
In addition, compared to heavy atoms, hydrogen offers a much cleaner theoretical interpretation, avoiding all the complexities of solving the relativistic many-body problem. 

To reiterate, hydrogen APV is uniquely sensitive to $\sin^2\theta _\mathrm{W}$ due the absence of the otherwise leading neutron contribution. This makes it an intriguing probe of the recently reported $7\sigma$ anomaly in the mass $M_W$ of W-boson~\cite{MW2022}. The $7\sigma$ anomaly translates into  W-boson being heavier by 0.1\%.
The implications of the W-boson mass for the hydrogen APV have been considered in Ref.~\cite{TraDer22}.  These authors pointed out that  ignoring the radiative corrections, $\sin^2 \theta_\mathrm{W} = 1 - M_W^2/M_Z^2$, so that the 0.1\% shift in $M_W$ is equivalent to a $\sim 0.5\%$ reduction in the  $\sin^2 \theta_\mathrm{W}$ value. This shift in the Weinberg angle is, however, strongly amplified  in the $C_{p}^{(1)}$ determination  due to $\sin ^2{\theta_\mathrm{W}} \approx 1/4$, see Eq.~\eqref{Eq:C1p}.  With the recent determination~\cite{MW2022} of the W-boson mass taken at face value, the proton spin-independent coupling $C_{p}^{(1)}$ value (or, equivalently, $\Qwp$, see Eq.~\eqref{Eq:Cp1-Qwp}) shifts by a sizable 3\%~\cite{TraDer22} from its previously accepted SM value. 

The proton weak charge $\Qwp$ has been deduced by the  Q$_{\rm weak}$ collaboration~\cite{Androic2018} from the measurements of parity-violating asymmetry in the scattering of polarized electrons on protons at the Jefferson Lab. Their reported value is
\begin{equation}
\Qwp = 0.0719(45) \,, \label{Eq:Qw-proton-JLAb} 
\end{equation}
which we will use as the nominal value in the rest of the paper. The Q$_{\rm weak}$ collaboration  result is
in agreement with the currently recommended SM value~\cite{Workman:2022ynf} 
\begin{equation}
\Qwp (\mathrm{SM}) = 0.0709(2) \,. \label{Eq:Qwp-SM}
\end{equation}
 Notice that 
the fractional uncertainty in the Q$_{\rm weak}$ collaboration result is 6\% and  can not decisively resolve the $ 3\%$ W-boson mass anomaly shift~\cite{TraDer22}. Determination of the proton weak charge at a $\sim 1\%$ level would be an important step in this direction.

Now we simplify the weak interaction Hamiltonian density $\mc{H}_\mathrm{PV,p}$, Eq.~\eqref{Eq:HPV_densityProton}, by taking the non-relativistic limit for the proton motion. Then $\bar p{\gamma ^\mu }p \rightarrow \psi_p^\dagger \psi_p \delta_{\mu,0}$ and $\bar p{\gamma ^\mu }{\gamma _5} p \rightarrow \psi_p^\dagger (\sigma_p)_i \psi_p \delta_{\mu,i}$, where  $\psi_p$ is the nonrelativistic proton wavefunction including spin and $\bm{\sigma}_p$ 
is the proton Pauli matrix. The former limiting expression gives rise to the proton spin independent contribution, while the latter -- to the spin-dependent terms. Recall that the nuclear density 
$\rho_p(\mathbf{r}) =\psi_p^\dagger \psi_p$, leading to the effective  operator in the electron sector,
\begin{equation}
 H_\mr{W} = - \frac{G_{\rm F}}{\sqrt{8}} \gamma_5 \Qwp \rho_p(r) \, . \label{Eq:HW-NSI}
\end{equation}
Here the proton distribution is normalized as $\int_0^\infty  4\pi r^2 \rho_p(r) dr = 1$. Similar simplification can be carried out for the proton spin dependent contributions. We quote the limiting result for the point-like proton and non-relativistic electron~\cite{Dunford1978},
\begin{align}
\label{Eq:VPV}
V_\mr{PV} =\frac{G_\mr{F}}{\sqrt{8} m_e c}\left\{ 
-C_{p}^{(1)}\left[\bm{\sigma} \cdot \bm{p}, \delta(\bm{r})\right]_{+}+C_{p}^{(2)}\left[\bm{\sigma}_{p} \cdot {\bm{p}}, \delta(\bm{r})\right]_{+}\right. \nonumber \\
\left. -i\left(C_{p}^{(2)} + C_{p}^{(3)}\right)\left[(\bm{\sigma} \times \bm{\sigma}_{p}) \cdot \bm{p}, \delta({\bm{r}})\right]_{-}
\right\} \,.
\end{align}
Here $\bm{p}$ is the electron linear momentum operator and Pauli matrices $\bm{\sigma}$
act on the electron spin. $[A,B]_{+}$ and $[A,B]_{-}$ denote anti-commutators and commutators, respectively. 
Notice the addition of the $C_p^{(2)}$ and $C_p^{(3)}$ constants in the last contribution to the potential. Since $C_p^{(3)}/C_p^{(2)} \approx \alpha/(2\pi) \ll 1$, we will neglect the anomalous weak moment contribution in our analysis. 

Previous efforts to measure parity violation in atomic hydrogen were initiated in the mid-1970s at the University of Michigan~\cite{Lewis1975,Dunford1978,Dunford1981c,Dunford1981,Levy1982,fehrenbach1993experimental}, Yale University~\cite{Hinds1977,Hinds1980,hinds1992parity}, and the University of Washington~\cite{adel1981,fortson1984atomic,wong1990search}, and continued through the early 1990s. In each of these programs, the investigators attempted to detect the microwave transition between two metastable hyperfine components of the $2s \, ^2\!S_{1/2}$ state.  This transition is nominally electric-dipole (E1) forbidden, but due to the weak interaction, it becomes slightly E1-allowed.  The group at Yale worked at zero magnetic field, while the Michigan and Washington groups applied a static magnetic field of $\sim$553 Gauss to Zeeman tune  components of the $2s \, ^2\!S_{1/2}$ and $2p \: ^2P_{1/2}$ states into near degeneracy in order to take advantage of the expected resonance-enhancement of the mixing. At this level-crossing, the measurement is primarily sensitive to the coupling coefficient $C_p^{(2)}$~\cite{Dunford1978}.  In each case, the investigators used an interference between the parity non-conserving interaction and a parity conserving interaction in order to amplify the former, as first suggested by Bouchiat and Bouchiat~\cite{bouchiat1974parity,bouchiat1975parity} in 1974-75. The Michigan group, which at its conclusion achieved a higher sensitivity than the Yale or Washington groups, reported a value of $C_p^{(2)} = 1.5 \pm 1.5_\mr{stat} \pm 22_\mr{syst}$~\cite{fehrenbach1993experimental}, where the first uncertainty is the statistical uncertainty, and the second is due to systematic factors.  The primary source of the systematic uncertainty resulted from the rotation of the microwave cavity, the so-called $\phi$-reversal in their measurement scheme. Note that this systematic uncertainty is greater than the SM value for $C_p^{(2)} = 0.0430$ by a factor of $\sim 500$.  

Recently, several papers have appeared in which the authors consider the prospects for renewed measurements in hydrogen~\cite{dunford1998pnc,Dunford2007a,Dunford2011,dekieviet2011longitudinal,Rasor2020}. 
Development of techniques for (i) generating beams of hydrogen of higher density, lower mean velocity, and decreased velocity dispersion~\cite{koch1999high,huber1999high,szczerba2000polarized,kolachevsky2004high,hogan2008slow,mikirtychyants2013polarized,Cooper2018}, and (ii) efficient optical excitation of the metastable $2s\, ^2\!S_{1/2}$ state~\cite{Cooper2018}, could lead to significant improvements, should they be employed in a new hydrogen PNC effort.


In this work, we reconsider the possibility of an {\em optical} APV measurement in hydrogen based on the $2s \, ^2\!S_{1/2} \rightarrow 3s \, ^2\!S_{1/2}$ transition at a wavelength of 656.46 nm, or the $2s \, ^2\!S_{1/2} \rightarrow 4s \, ^2\!S_{1/2}$ transition at a wavelength of 486.27 nm.  An APV measurement on the former was considered previously by Lewis and Williams~\cite{Lewis1975}, and more recently by Rasor and Yost~\cite{Rasor2020}.   In Sec.~\ref{Sec:theory}, we calculate the matrix elements of weak interaction Hamiltonian between the $3s$ and $3p_{1/2}$ states and between the $4s$ and $4p_{1/2}$ states, using relativistic wavefunctions, and including the effect of the finite size of the nucleus. This theoretical approach is more sophisticated compared to the earlier literature, in part to match the accuracy of the SM value of the proton weak charge, Eq.~\eqref{Eq:Qwp-SM}.  In Sec.~\ref{sec:Exp}, we use these results to determine the magnitude of the parity violating amplitude for these two transitions, identify the Zeeman crossings that could be useful in extracting the coupling coefficients $C_{p}^{(1)}$ and $C_{p}^{(2)}$, and briefly outline a possible measurement scheme based on interfering coherent interactions, using two-photon absorption as the local oscillator signal.  While we do not explicitly consider deuterium in this work, similar measurements in this system offer the possibility of determining $C_{n}^{(1)}$ and $C_{n}^{(2)}$ as well, as has been recognized previously~\cite{Dunford2007a}.

\section{Theory}
\label{Sec:theory}

In the previous section, we reviewed the  theory of electroweak interactions as it relates to APV. Here we focus on evaluating weak-interaction matrix elements for hydrogen and refine the previous literature analysis by using fully relativistic treatment and a realistic nuclear distribution.

The weak interaction~\eqref{Eq:HW-NSI}  is a pseudo scalar operator, mixing atomic states of opposite parity and of the same total angular momenta. Explicitly, the parity-mixed states $\overline{|n \ell_{j} m_j \rangle}$  can be expanded over the conventional parity-proper states ${|n \ell_{j} m_j \rangle}$ as  
\begin{equation}
\overline{|ns_{1/2}  m_j\rangle}  = |ns_{1/2} m_j \rangle +  \sum_{n'=2}^{\infty} \eta_{nn'} |n'p_{1/2} m_j\rangle \, ,
\end{equation}
where the mixing coefficients $\eta_{n n'}$ can be computed perturbatively,
\begin{equation}
\label{PNC_Mixing_Coefficient}
\eta_{n n'} = \frac{\langle n'p_{1/2} m_j | H_{\mr{W}} | ns_{1/2} m_j \rangle}{E_{ns} - E_{n'p_{1/2}} + \frac{i}{2}  (\Gamma_{n'p_{1/2}} - \Gamma_{ns})} \, .
\end{equation}
Here $\Gamma$'s are the (radiative) level widths. In hydrogen-like systems, the dominant admixtures for the excited $ns_{1/2}$ states comes from  $n'=n$  due to level degeneracies. We remind the reader that for a point-like nucleus the $ns_{1/2}$ and $np_{1/2}$ levels remain degenerate even in the fully relativistic treatment; only the QED and finite nuclear size corrections lift this degeneracy. We also assume that there are no external fields applied and discard the nuclear-spin dependent effects; these will be addressed in Sec.~\ref{sec:Exp}.

The relevant  matrix element of the weak interaction~\eqref{Eq:HW-NSI}  reads
\begin{align}
  \label{PNC_Matrix_Element_R}
\langle np_{1/2} m_j |H_{\mr{W}}| ns_{1/2} m_j \rangle &= -i\frac{G_{\mathrm{F}}}{\sqrt{8}} \Qwp \times \nonumber\\ 
& \int_{0}^{\infty} dr \rho_{p}(r) \left[F^*_{np_{1/2}}(r)G_{ns}(r) -  G^*_{np_{1/2}}(r)F_{ns}(r) \right] \, ,
\end{align}
where $F$ and $G$ are the large and small radial components of relativistic bi-spinors~\cite{WRJBook}. The matrix element does not depend on the value of magnetic quantum number $m_j$, reflecting the pseudo-scalar nature of the interaction.
It is worth emphasizing that the expression is fully relativistic in electron degrees of freedom and involves proton distribution $\rho_{p}(r)$. The same nuclear charge distribution also affects the wavefunctions. Below we explore the effects of various approximations in computing weak-interaction matrix element~\eqref{PNC_Matrix_Element_R}.  We start by using a point-like nucleus $\rho_{p}(r) \propto \delta(r)$ both in computation of wavefunctions and while evaluating the matrix element. This leads to a fully-relativistic result that recovers literature expressions in the non-relativistic limit.
Then we proceed to computations with a more realistic nuclear distribution.

With relativistic orbitals~\cite{Greiner2000} for H-like ions and point-like nucleus, we obtain the matrix element
\begin{align}
\label{PNCmatrixelement_PN_R}
\langle np_{1/2} m_j |H_{\text{W}}| ns_{1/2} m_j \rangle =& i \frac{\sqrt{2}}{2\pi} G_{\text{F}} \Qwp Z^4\alpha \frac{\Gamma(2\gamma+1+n_r)}{\Gamma^2(2\gamma+1)} \times \\
&\left[ \frac{n-1}{n_r! N^5 \sqrt{N^2-1}} \right] \exp\left[ (Z\alpha)^2 \ln\frac{1}{\rho} \right] \, . \nonumber 
\end{align}
{Here, $\alpha$ is the fine-structure constant, nuclear charge $Z=1$ for hydrogen, $\gamma = \sqrt{1- (\alpha Z)^2}$, the radial quantum number
$n_r=n-1$,
$N=[n^2-2n_r(1-\gamma)]^{1/2}$, and $\rho = 2 \lambda r$ with $\lambda=Z/N$. 
}
It should be emphasized that the analytical relativistic wavefunctions used in arriving at Eq.~\eqref{PNCmatrixelement_PN_R} were obtained using the point-like nucleus. In other words, the effect of the nuclear density distribution on the wavefunctions has been neglected. A direct consequence of this assumption is the singularity in the $j=1/2$ relativistic wavefunctions, leading to the appearance of the factor $\exp\left[ (Z\alpha)^2 \ln\frac{1}{\rho} \right] \propto r^{-(\alpha Z)^2}$ in Eq.~\eqref{PNCmatrixelement_PN_R}. For hydrogen, this is a mildly-diverging factor in the $r\rightarrow 0$ limit. Varying $r= 10^{-6}$ a.u. to $r= 10^{-5}$ a.u. ($\approx 2\,\mr{fm}$) modifies this factor by just $\sim 0.01\%$ for $Z=1$. A natural scale for $r$  is the proton charge radius, $r_p = 0.84 \, \mathrm{fm} \approx 1.59 \times 10^{-5} \, \text{a.u.}$.  Moreover, the divergence 
 is easily cured using a more realistic nuclear charge distribution while evaluating the integral in~\eqref{Eq:HW-NSI}, see below. Also the divergence goes away if the wavefunctions are computed using realistic nuclear charge distributions, see, e.g., Ref.~\cite{Dra05AMOHandBook}.

The non-relativistic limit of the PNC matrix element~\eqref{PNCmatrixelement_PN_R} reads
\begin{equation}
\label{PNCmatrixelement_PN_NR}
\langle np_{1/2} m_j |H_{\text{W}}| ns_{1/2} m_j \rangle \approx i \frac{\sqrt{2}}{8\pi} G_F \Qwp Z^4\alpha \left[ \frac{\sqrt{n^2-1}}{n^4} \right] \,,
\end{equation}
where we took the  $\alpha Z \rightarrow 0$ limit, so that $\gamma \approx 1$, $N \approx n$ and $\exp\left[ (\alpha Z)^2 \ln\frac{1}{\rho} \right] \approx 1$. This limiting expression recovers the literature results~\cite{Feinberg1974,Dunford1978,Drake1988}. This expression implies the $1/n^3$ scaling for PNC matrix elements.

We calculated PNC matrix elements for the $2s_{1/2}-2p_{1/2}$, $3s_{1/2}-3p_{1/2}$ and $4s_{1/2}-4p_{1/2}$ pairs of states in different approximations, and collected these results in Table~\ref{PNC_Matrix_Element_Table}.
The non-relativistic (NR) PNC matrix elements in the point-like nucleus (PN) approximation, labeled as PN-NR, were obtained with Eq.~(\ref{PNCmatrixelement_PN_NR}) and their relativistic counter-parts PN-R --- with Eq.~(\ref{PNCmatrixelement_PN_R}), where we used $r=10^{-6}$ a.u.
At the next step we  explore the effect of using the finite-sized nucleus (FN) instead of the point-like nucleus. 
The nuclear distribution $\rho_{p}(r)$ can modify the PNC matrix elements in two distinct ways: (i) $\rho_{p}(r)$ appears directly in the integral~\eqref{PNC_Matrix_Element_R} and (ii) it can modify the atomic wavefunctions in the nuclear region. 
 To model finite-sized nuclear distribution we employed the Fermi distribution
\begin{equation}\label{FMNCD}
\rho_p(r) = \frac{\rho_0}{1+\exp[(r-c)/a]} \,, 
\end{equation}
where $c$ is the 50\% falloff radius of the charge distribution, $a$ is related to the 90\% - 10\% falloff distance $t$ as $t=4 \ln3~a$, and $\rho_0$ is the normalization constant. For the proton we used {$c=0.4$~fm and $a=0.18$~fm to fit the proton root-mean-square radius ($r_p \sim 0.855$~fm)}.

FN-NR and FN-R entries in Table~\ref{PNC_Matrix_Element_Table} show the direct effect of the proton charge distribution $\rho_p(r)$ as in these approximations we used the analytic non-relativistic and relativistic wavefunctions for a point-like nucleus.   In the FN-NR approximation, 
we used the Pauli approximation in Eq.~\eqref{Eq:HW-NSI} to arrive at the PNC operator~\cite{Khriplovich1991} 
\begin{equation}
\label{NR_PNC}
H^\mr{NR}_\mr{W} = \frac{1}{2c} \frac{G_F}{\sqrt{8}} \mathcal{Q}_{W} [ ({\bm \sigma} \cdot {\bf p})^{\dagger} \rho_{p}(r) + \rho_{p}(r) ({\bm \sigma} \cdot {\bf p})] \,, 
\end{equation}
where the dagger form of the operator $({\bm \sigma} \cdot {\bf p})$ acts on the bra when computing matrix elements.

Finally, in addition to the direct effect in Eq.~\eqref{PNC_Matrix_Element_R}, the proton charge distribution  modifies atomic wave functions, in turn shifting the values of the PNC matrix elements.
To account for this ``indirect'' effect, we  solved the  radial Dirac equation numerically 
with the Fermi distribution~\eqref{FMNCD}.
As can be seen from the last column (marked Full) of Table~\ref{PNC_Matrix_Element_Table}, the influence of the finite-size proton on the $2s-2p_{1/2}$ PNC matrix element is negligible due to the smallness of proton. This indirect effect can be ignored for the $3s-3p_{1/2}$ and $4s-4p_{1/2}$ PNC matrix elements since the atomic wavefunctions with larger principal quantum numbers are pushed further away from the origin.

Examining  numerical results in Table~\ref{PNC_Matrix_Element_Table},  we find that the dominant effect on the PNC matrix elements is due to using relativistic wavefunctions, rather than the finite nuclear size effect. The matrix element variations across different approximations for nuclear distributions are just $\sim 0.02\%$, which is well below the current $0.3\%$ uncertainty in the SM proton weak charge value~\eqref{Eq:Qwp-SM}. This statement must be contrasted with APV in $^{133}$Cs, where the uncertainty in the nuclear distribution is anticipated to become a limiting factor in extracting  precise value of the nuclear weak charge~\cite{BroDerFla09,Der01.Breit.NeutronSkin}. 

\begin{table}[!ht]
\caption{\label{PNC_Matrix_Element_Table} Weak interaction matrix elements $-i \langle np_{1/2} m_j |H_\mr{W}| ns_{1/2} m_j \rangle$  (in a.u.) multiplied by $10^{21}$. PN: point-like nuclear  distribution in the PNC Hamiltonian; FN: Fermi nuclear charge distribution in the PNC Hamiltonian; NR: non-relativistic wave functions obtained with the point-like nuclear potential; R: relativistic wave functions obtained with the point-like nuclear potential. The column marked ``Full'' includes full numerical relativistic treatment where we used Fermi distribution in computing both the wavefunctions and the PNC matrix element~\eqref{PNC_Matrix_Element_R}.}
\begin{tabular}{cccccccccccccccc}
\hline 
\hline
        &   PN-NR    &   FN-NR    &    PN-R     &    FN-R     &   Full      \\
\hline
$2s-2p_{1/2}$   &   71.03    &   71.03    &    71.09    &    71.07    &           71.07                \\  
$3s-3p_{1/2}$   &   22.91    &   22.91    &    22.93    &    22.92    &                                \\
$4s-4p_{1/2}$   &   9.927    &   9.927    &    9.935    &    9.930    &                                \\
\hline
\hline        
\end{tabular}
\end{table}

With the PNC matrix elements computed, now we compile the mixing coefficients $-i \eta_{nn}$,  Eq.~\eqref{PNC_Mixing_Coefficient}, see Table~\ref{Table:PNC_Mixing_Coefficient}.  We used accurate energies of the $ns_{1/2}$ and $np_{1/2}$ states from Ref.~\cite{Kramida2010,horbatsch2016tabulation,Bezginov2019,Fabjan1971,fleurbaey2018new}. We estimated radiative  widths $\Gamma$ by summing spontaneous emission rates~\cite{NIST} over all the possible E1 decay channels. The relevant atomic data are collected in Table~\ref{Table:Hproperties}. 

We find that all the tabulated PNC mixing coefficients are comparable, even though the PNC matrix elements scale as $1/n^3$. This is attributed to the decreasing energy splittings with increasing principal quantum number.  It is worth reemphasizing that our calculations in this section assume the absence of magnetic fields; Zeeman level crossings are discussed in Sec.~\ref{sec:Exp}. 

\begin{table}[!ht]
\caption{\label{Table:PNC_Mixing_Coefficient} PNC mixing coefficients $-i \eta_{nn}$, Eq.~\eqref{PNC_Mixing_Coefficient}, multiplied by a factor of $10^{13}$. $\Delta E_n = E_{ns} - E_{np_{1/2}}$ is the energy splitting, and $\Delta \Gamma = \Gamma_{np_{1/2}} - \Gamma_{ns}$ is the effective energy width.  
}  
\begin{ruledtabular}
\begin{tabular}{cccccccccccccccc}
        &   $\Delta E_n$~(MHz)  &  $\Delta \Gamma$~(MHz)  &   PN-NR    &     FN-NR    &     PN-R    &     FN-R     &    Others                      \\      
\hline
$2s-2p_{1/2}$   &       1057.847         &       99.709               &   4.408    &     4.408    &    4.412    &     4.410    &   
4.3\footnote{\citet{Moskalev1974}; \citet{Lewis1975} } ; 
3.6\footnote{\citet{Dunford1978}}       \\  
$3s-3p_{1/2}$   &       314.818           &       29.187               &   4.779    &     4.778    &    4.782    &     4.780    &                                \\
$4s-4p_{1/2}$   &       133.2            &       12.231               &   4.893    &     4.893    &    4.897    &     4.895    &
\end{tabular} 
\end{ruledtabular}
\end{table}

\begin{table}[!ht]
\caption{\label{Table:Hproperties} 
Properties of the hydrogen atom used in our numerical evaluations. $\Delta E_{\text{Lamb}}$ is the energy difference (Lamb shift) between the $ns$ and $np_{1/2}$ states and 
 $\Gamma$ are radiative level widths at zero magnetic and electric fields. 
} 
\begin{tabular}{ccccccccccccccc}
\hline 
\hline
State      &  $\Delta E_{\text{Lamb}}$ (MHz)  &   $\Gamma$ (MHz)   \\
\hline
$2s_{1/2}$ &    1057.847                     &          0              \\
$3s_{1/2}$ &     314.82                      &         1.0050          \\
$4s_{1/2}$ &     133.2                        &         0.70251      \\[1.5ex] 
$2p_{1/2}$ &                                 &         99.709          \\
$3p_{1/2}$ &                                 &         30.192          \\
$4p_{1/2}$ &                                 &         12.934          \\
\hline
\end{tabular}
\end{table}

The PNC amplitude for the $n's-ns$ transition~\cite{Porsev2009} can be simplified to 
\begin{align}
E_{\textrm{PNC}} =& \overline{\langle ns_{1/2} m_j} | D_z | \overline{n's_{1/2} m_j \rangle} \nonumber\\
                =&   \frac{1}{3} \left[ \sum_{\mu} \eta^*_{n \mu} ( \mu p_{1/2} ||r|| n's ) + \sum_{\nu} \eta_{n' \nu} ( ns ||r|| \nu p_{1/2} ) \right] \label{PNC_amplitude}\, , 
\end{align}
where 
$( \mu p_{1/2} ||r|| n's )$ and $( ns ||r|| \nu p_{1/2} )$ are the radial integrals of the electric dipole operator. Notice that the overall phase of the PNC amplitude is not fixed as it depends on the relative phases of the two $s_{1/2}$ states.

PNC amplitudes in hydrogen are dominated by the mixing between the near-degenerate energy levels. Then the $2s-3s$ and $2s-4s$ PNC amplitudes evaluate to
\begin{align}
\label{Eq:PNC_amplitude_2s_4s}
E_{\textrm{PNC}}(2s-3s) = \frac{1}{3} \left[ \eta^*_{33} ( 3p_{1/2} ||r|| 2s ) + \eta_{22} ( 3s ||r|| 2p_{1/2} ) \right] =  3.504 i \times 10^{-13}~\textrm{a.u.} \, , \nonumber \\
E_{\textrm{PNC}}(2s-4s) = \frac{1}{3} \left[ \eta^*_{44} ( 4p_{1/2} ||r|| 2s ) + \eta_{22} ( 4s ||r|| 2p_{1/2} ) \right] =  1.531 i \times 10^{-13}~\textrm{a.u.}\, ,
\end{align}
where we used the FN-R mixing coefficients from Table~\ref{Table:PNC_Mixing_Coefficient}.
In addition, the dipole matrix elements were obtained with the relativistic analytic wave functions.
Note that the $2s-3s$ PNC amplitude is three times larger than the $2s-4s$ one, mainly because of its larger dipole matrix element. The latter is just a little smaller than that for the $1s-2s$ transition,  i.e. $E_{\textrm{PNC}} (1s-2s) =  1.897i \times 10^{-13}~\textrm{a.u.}$. 

\section{Possible experimental realization}\label{sec:Exp}
\begin{figure}
    \centering
    \parbox{6.5cm}{
     \includegraphics[width=5.5cm]{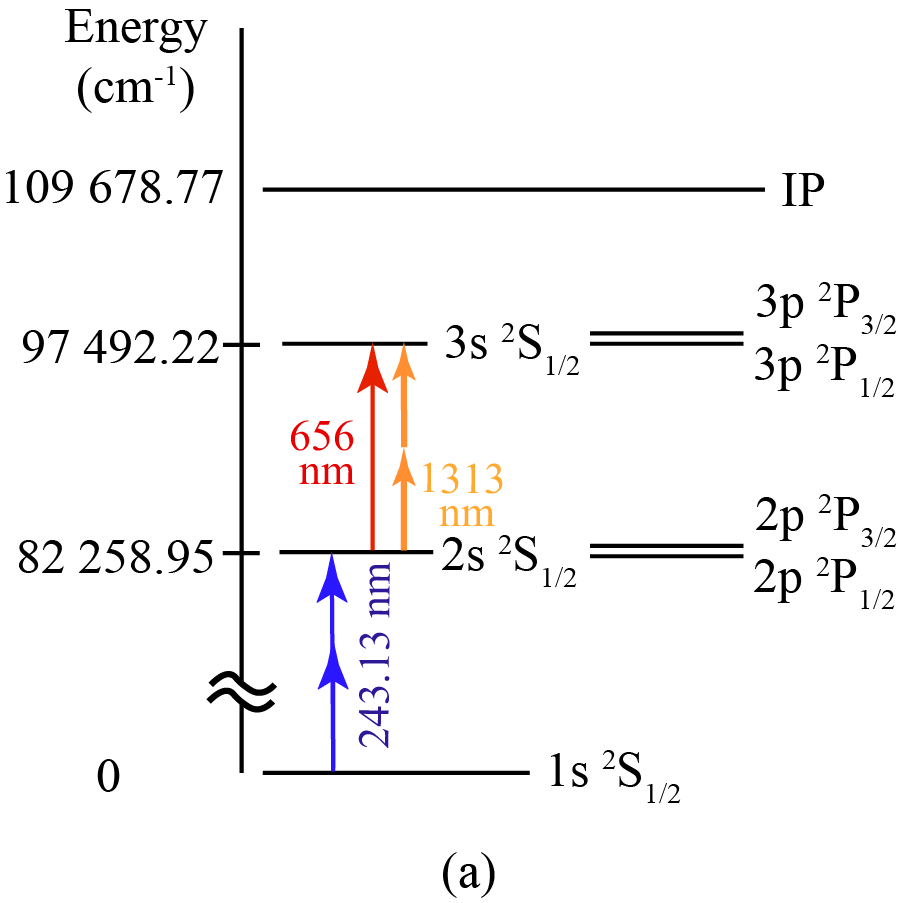}  \vspace{0.2in} \includegraphics[width=6.7cm]{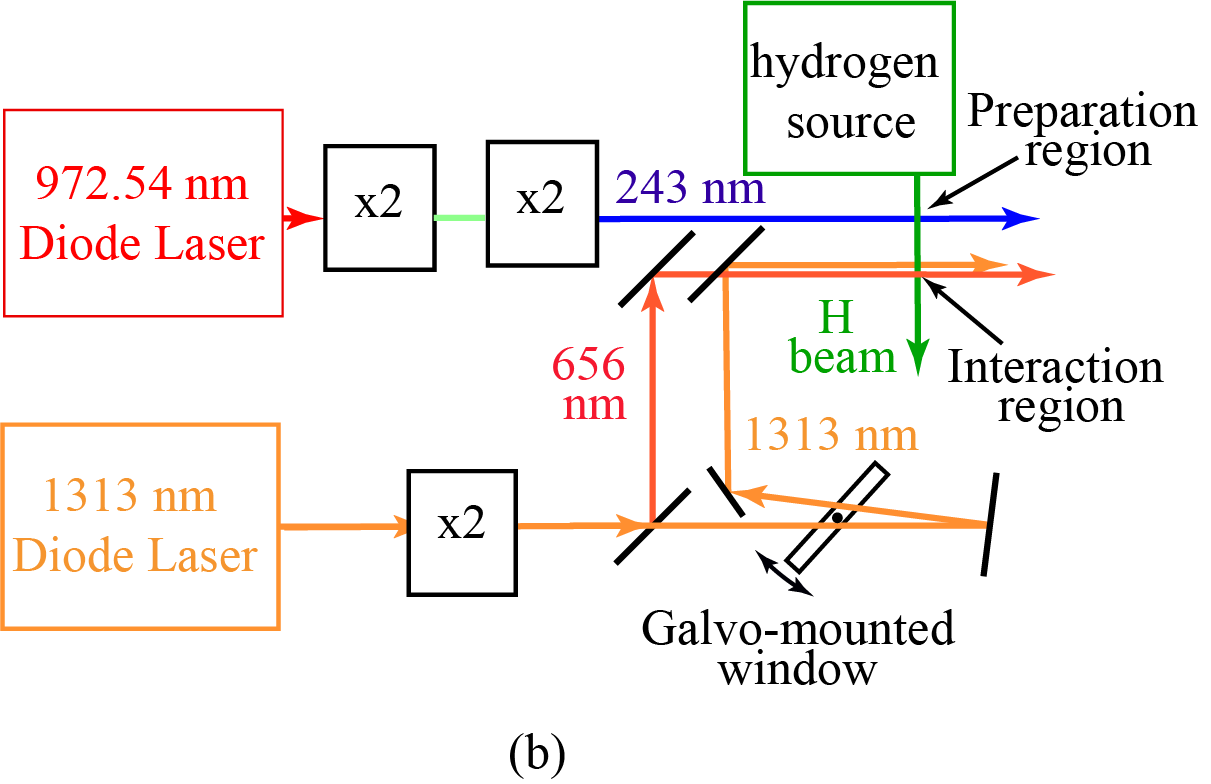}} 
 \parbox{8cm}{\vspace{-0.2in}\includegraphics[width=8cm]{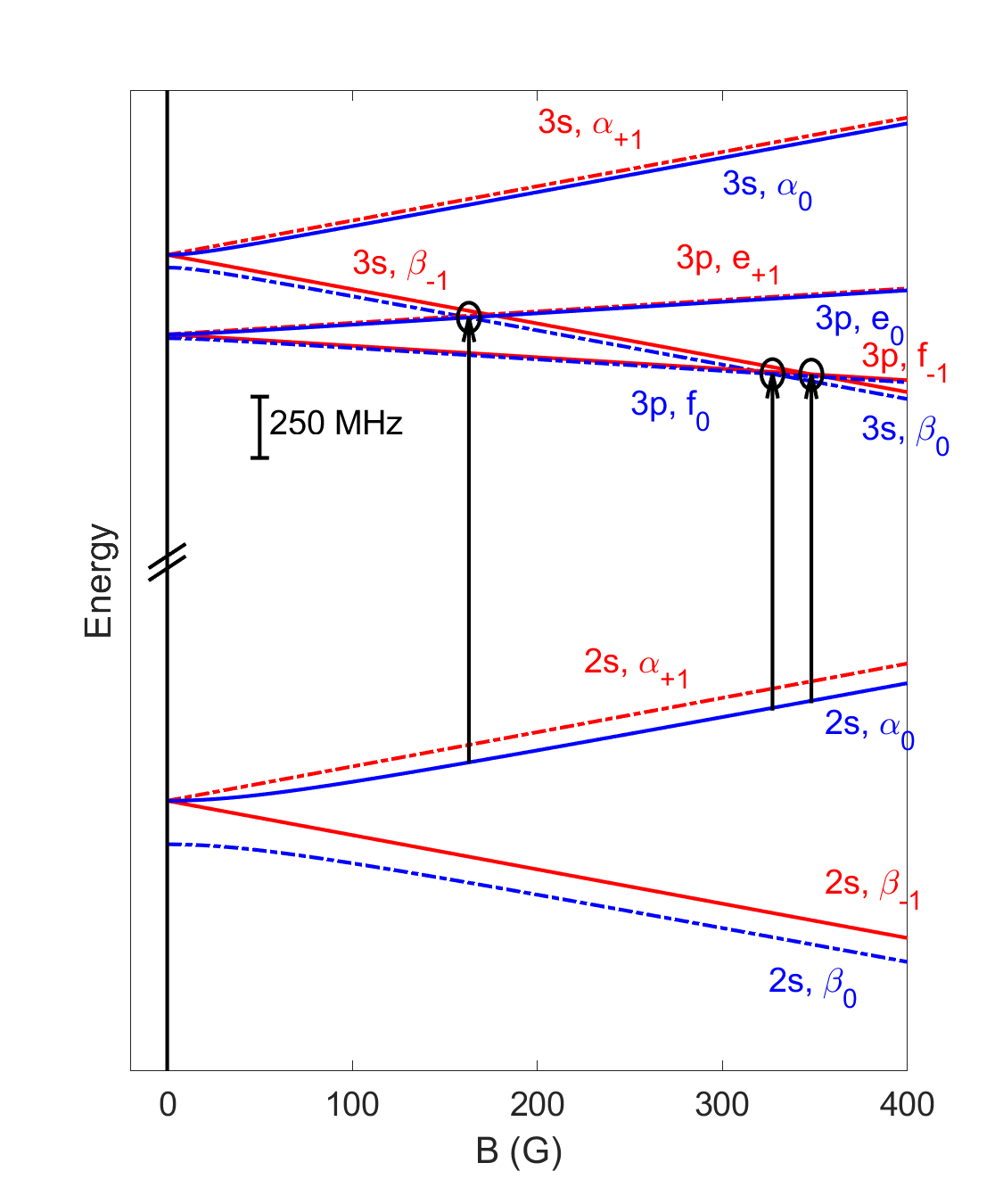}  \hspace{1in} (c)}
    \caption{(a) An energy level diagram showing the relevant levels of atomic hydrogen. (b) Experimental geometry for a coherent control measurement of $E_{\textrm{PNC}}$ on the $2s \: ^2S_{1/2} \rightarrow 3s \: ^2S_{1/2}$ transition. (c) An energy level diagram showing the level crossings induced by the Zeeman shift.  Three level crossings between the $3s \: ^2S_{1/2}$ and $3p \: ^2P_{1/2}$ are of interest in this work:  $\beta_0 - e_0$ at B = 163 G, $\beta_0 - f_0$ at B = 327 G, and $\beta_{-1} - f_{-1}$ at B = 348 G.  The optical transition from the $2s \: ^2S_{1/2}, \alpha_0$ level will be driven by the two-color coherent laser field at 1313 nm and 656 nm simultaneously.} 
    \label{fig:exp}
\end{figure}

We turn now to consider how this PNC moment might be measured in a laboratory. For this we suggest a two-color coherent-control process, as has been demonstrated in Refs.~\cite{AntypasE2013,AntypasE2014}. Rasor and Yost~\cite{Rasor2020} have considered a similar scheme.  This all-optical approach is similar in some regards to the microwave measurements of the Michigan~\cite{Lewis1975,Dunford1978,Dunford1981c,Dunford1981,Levy1982,fehrenbach1993experimental}, Yale~\cite{Hinds1977,Hinds1980,hinds1992parity}, and  Washington~\cite{adel1981,fortson1984atomic,wong1990search} hydrogen PNC experiments.  
To measure the $\langle 3p_{1/2} m_j |H_{W}| 3s_{1/2} m_j \rangle$ matrix elements, we suggest generating a slow, high-density atomic hydrogen beam, exciting the ground state atoms to the metastable $2s \: ^2S_{1/2}, \: \alpha_0$ state via two-photon excitation with 243.1 nm light, and driving the $2s \: ^2S_{1/2}, \: \alpha_0 \rightarrow 3s \: ^2S_{1/2},  \: \beta_0$ transition using three distinct interfering interactions ($\alpha_0$ and $\beta_0$ are two hyperfine components of the $ns \: ^2S_{1/2}$ state, using notation introduced by Lamb and Retherford~\cite{lamb1950fine}.) 
This transition is electric dipole (E1) forbidden, but is active through Stark mixing when a static electric field is applied, as well as through the weak-interaction mixing. The wavelength of the laser beam that is resonant with this transition is 656.5 nm.  See Fig.~\ref{fig:exp}(a) for a partial energy level diagram of hydrogen.
The transition is also active as a two-photon transition, driven by a field at a wavelength of $\lambda = 1313$ nm.  (The two-photon interaction can also be driven with two unequal-frequency beams, providing flexibility of the selection rules for the transition, but also introducing experimental challenges. We will consider only the simpler form for now.) 
When the three interactions are driven concurrently, the net transition rate scales as the square of the sum of the individual amplitudes,
\begin{equation}
    \mathcal{R} = \left| A_{2\gamma} + A_{St} + A_{PNC} \right|^2,
\end{equation}
where $A_{2\gamma}$, $A_{St}$, and $A_{PNC}$ are the amplitudes for the two-photon, Stark-induced, and PNC-induced interactions, respectively. 
{In the coherent control technique, only modest electric fields are required, such that the maximum Stark amplitude $|A_{St}|$ is less than $\sim5 \times A_{PNC}$, and since $A_{PNC}$ is very weak,}
a two-photon amplitude that is much greater than $A_{St}$ or $A_{PNC}$ is easily achieved.  In this limit,
\begin{equation}
    \mathcal{R} \approx \left| A_{2\gamma} \right|^2 + A_{2\gamma}^{\ast} (A_{St} + A_{PNC}) + c.c.,
\end{equation}
where we have omitted the small $|A_{St} + A_{PNC}|^2$ term.  In this expression, $\left| A_{2\gamma} \right|^2$ represents the two-photon absorption rate by itself, while the cross term arises from the interference between the weak amplitudes ($A_{St}$ and $A_{PNC}$) and the strong two-photon amplitude.  Coherence between the 1313 nm field and the 656 nm field is assured when the latter is generated from the former through a second-harmonic generation process, which is inherently coherent. We illustrate a layout of laser sources in Fig.~\ref{fig:exp}(b).  A laser beam of wavelength 243 nm, generated through two stages of frequency doubling of the output of a 972.54 nm high-power laser diode, can be used to excite the $F=1$ ($\alpha_0$) component of the metastable $2s \,^2\!S_{1/2}$ state.  A second laser source, of wavelength 1313 nm, can be used to excite the $2s\, ^2\!S_{1/2} \rightarrow 3s\, ^2\!S_{1/2}$ transition via two-photon absorption, or, after frequency doubling to generate a 656 nm beam, through a linear interaction.  The $2s\, ^2\!S_{1/2}$ state is long lived, so loss of population as the atoms travel from the preparation region to the region where they interact with the two-color field (656 nm plus 1313 nm) is minimal.  Upon phase shifting the 1313 nm beam, the interference between the linear excitation and the two-photon excitation of the $3s ^2S_{1/2}$ state can be modulated, and the amplitude of this modulation, normalized by the two-photon rate alone, is
\begin{equation}
    2 \left\{\frac{A_{St} + A_{PNC}}{A_{2\gamma}} \right\}.
\end{equation}
Since the amplitudes $A_{St}$ and $A_{PNC}$ are $90^{\circ}$ out of phase with one another, the amplitude of the interference term scales as 
\begin{equation}
    2 \left\{\frac{\sqrt{|A_{St}|^2 + |A_{PNC}|^2}}{A_{2\gamma}} \right\}
\end{equation}
and measurement of this amplitude as a function of the applied static electric field can be used to determine the ratio $A_{PNC}/A_{St}$.

{By applying a magnetic field to the atoms in the interaction region, the energy levels of the $3s$ and $3p_{1/2}$ states, and similarly the $4s$ and $4p_{1/2}$ states, can be made to cross. We illustrate the level crossings of the $3s$ and $3p_{1/2}$ states in Fig.~\ref{fig:exp}(c). By bringing the levels into degeneracy, the admixture of states caused by the weak interaction can be enhanced by a factor of $\Gamma/\Delta E$, where $\Gamma$ is radiative width of the state, and $\Delta E$ is the energy splitting at zero magnetic field.  This enhancement factor is $\sim$21 for the $3s$ - $3p_{1/2}$  and the $4s$ - $4p_{1/2}$ manifolds, very similar to that of the $2s$ - $2p_{1/2}$ states. Bringing levels into degeneracy also allows for selection of specific combinations of coupling constants.  We have evaluated the Zeeman mixing of states, and the weak force interaction elements, and find very similar results for the $3s$ - $3p_{1/2}$ and $4s$ - $4p_{1/2}$ level crossings as were previously reported in the $2s$ - $2p_{1/2}$ system.  We compile these in Table~\ref{table:LevelCrossings}.  In this table, we include the magnetic field and the dependence of the weak interaction term on the coupling constants $C_{p}^{(1)}$ and {$C_{p}^{(2)}$}, which we determine by evaluating matrix elements of Eq.~\eqref{Eq:VPV}~\cite{Dunford1978}  
\begin{equation}
   \langle n s, m_J^{\prime} m_I^{\prime}|V_\mr{PV}| np_{1/2}, m_J m_I \rangle = +i \overline{V} \langle m_J^{\prime} m_I^{\prime}|\left[ -C_{p}^{(1)} + 2 C_{p}^{(2)} \bm{\sigma} \cdot \mathbf{I} \right] |  m_J m_I \rangle \,. 
\end{equation}
(We have omitted the third coupling constant $C_{p}^{(3)}$, which is expected to be much smaller than $C_{p}^{(2)}$ per discussion in Sec.~\ref{Sec:Intro}.) 
In this expression, $m_J$ and $m_J^{\prime}$ are the electronic and $m_I$ and $m_I^{\prime}$ are the nuclear magnetic quantum numbers. $\bm{\sigma}$ is the Pauli spin operator for the electronic angular momentum, and $\mathbf{I}$ is the nuclear spin operator. The scaling factor $\overline{V} = (3 G_\mathrm{F} /8 \pi \sqrt{2}) R_{n0}(0) R^{\prime}_{n1}(0) (\hbar/m_e c )$ is as defined in Eq.~(26) of Ref.~\cite{Dunford1978}, where $R_{n\ell}(0)$ is the radial non-relativistic wavefunction at the origin and prime denotes a derivative with respect to $r$. Evaluation of these terms with wavefunctions tabulated, e.g., in Ref.~\cite{Dra05AMOHandBook} leads to 
\begin{equation}\label{eq:zeta}
   \langle n s, m_J^{\prime} m_I^{\prime}|V_\mr{PV}| np_{1/2}, m_J m_I \rangle = +i \overline{V} \left\{ \zeta_1 C_{p}^{(1)} + \zeta_2 C_{p}^{(2)} \right\} .  
\end{equation}
Since in this section we focus on the experimental feasibility, we used the approximation with a point-like nucleus and non-relativistic electronic wave-functions, which are sufficient for this goal, unless the experimental uncertainties reach the $<0.1\%$ level.
We have listed the coefficients $\zeta_1$ and $\zeta_2$ for each level crossing in Table \ref{table:LevelCrossings}.
\begin{table}[]
    \centering
    \begin{tabular}{c|c|c|c}
   \rule{0.1in}{0in} matrix element \rule{0.1in}{0in} & \rule{0.3in}{0in} $\zeta_1$ \rule{0.3in}{0in}  &  \rule{0.3in}{0in} $\zeta_2$  \rule{0.3in}{0in} & \rule{0.3in}{0in} $B$ (G)\rule{0.3in}{0in} \\  \hline \hline
 \multicolumn{1}{l|}{\underline{$2s$ - $2p_{1/2}$}} & & &  \\
$\langle \beta_{0}| V_\mr{PV} | e_0 \rangle$  & 0.0002  &  -1.9858  &  552.14 \\
$\langle \beta_{0}| V_\mr{PV} | f_0 \rangle$  & -0.9974  &  -0.8889  &  1157.39 \\
$\langle \beta_{-1}| V_\mr{PV} | f_{-1} \rangle$  & 0.9970  &  -0.9970  &  1230.53 \\
 \multicolumn{1}{l|}{\underline{$3s$ - $3p_{1/2}$}} & & &  \\
$\langle \beta_{0}| V_\mr{PV} | e_0 \rangle$  & 0.0006  &  -1.9854  &  164.39 \\
$\langle \beta_{0}| V_\mr{PV} | f_0 \rangle$  & -0.9973  &  -0.8894  &  344.66 \\
$\langle \beta_{-1}| V_\mr{PV} | f_{-1} \rangle$  & 0.9969  &  -0.9969  &  366.34 \\
 \multicolumn{1}{l|}{\underline{$4s$ - $4p_{1/2}$}} & & &  \\
$\langle \beta_{0}| V_\mr{PV} | e_0 \rangle$  & 0.0013  &  -1.9846  &  69.48 \\
$\langle \beta_{0}| V_\mr{PV} | f_0 \rangle$  & -0.9973  &  -0.8896  &  145.69 \\
$\langle \beta_{-1}| V_\mr{PV} | f_{-1} \rangle$  & 0.9969  &  -0.9969  &  154.84 \\ \hline
    \end{tabular}
    \caption{The coefficients $\zeta_1$ and $\zeta_2$ of Eq.~(\ref{eq:zeta}) yielding the relative contribution of $C_{p}^{(1)}$ and $C_{p}^{(2)}$ to $\langle n s, m_J^{\prime} m_I^{\prime}|V_\mr{PV}| np_{1/2}, m_J m_I \rangle$ at the different level crossings.  Results for the different states $n=$2, 3, and 4 are quite similar to one another.  In the fourth column, we list the magnetic field of the level crossings.}
    \label{table:LevelCrossings}
\end{table}
The values of $\zeta_1$ and $\zeta_2$ for the $2s$ - $2p_{1/2}$ level crossings are close to, but not quite the same, as those reported by Dunford, Lewis, and Williams~\cite{Dunford1978}. Notice that the weak matrix element near the $\beta_{0}$ - $e_0$ level crossing is relatively insensitive to the $C_{p}^{(1)}$ coupling constant for $n=2$, 3, and 4, and this level crossing provides a hopeful avenue for a precision determination of $C_{p}^{(2)}$.
In the absence of a magnetic field, a similar analysis leads to the matrix element 
\begin{equation}
   \langle n s \beta_0|V_\mr{PV}| np_{1/2} \rangle = +i \overline{V} \left\{ - C_{p}^{(1)} +  C_{p}^{(2)} \right\} \,.  
\end{equation}}

{Systematic errors are known to play a critical role in PNC measurements, and careful attention to these effects is required to reduce their impact.  Stray electric fields, which can strongly mix the $ns_{1/2}$ and $np_j$ states, are expected to be more critical for $n=3$ or $4$ in the optical measurement than for $n=2$ in the rf measurements, due to the larger electric dipole matrix elements $\langle ns_{1/2} || r || np_j \rangle$ for $n=3$ or $4$, and the smaller Lamb shift. On the other hand, the optical coherent control method eliminates the rf cavity, and requires only modest static electric fields, allowing large spacing of the parallel field plates used to control the electric field. The larger distance between the hydrogen atoms and these conductors reduces the effect of surface contaminants and patch effects, which could introduce uncontrollable electric fields in the interaction region~\cite{speake1996forces}. Still, control of the electric field is expected to be a challenge in this measurement.  Using a perturbative analysis, we estimate that for a dc electric field of magnitude $E_0 \sim 5 \: \mu$V/cm, the Stark-induced amplitude $A_{St}$ matches the PNC amplitude, $A_{PNC}$.  While this critical $E_0$ is $\sim$100 times greater than that for the $2s \: \beta_0 \rightarrow 3s \: \beta_0$ transition suggested in Ref.~\cite{Rasor2020}, achieving this level of control will require development of new laboratory techniques.

Reduction of systematic contributions to the PNC measurement, relative to the single-field rf measurement, are also expected from the method used for phase control. In the coherent control, optical measurement, shifting the laser field takes place external to the interaction region.  In addition, the uniformity and tight control of linear polarization of laser beams, and the state selectivity of the two-photon interaction (that is, only transitions coupling the same initial and final states can interfere with the two-photon interaction) serve to further reduce systematic effects. Tight control of laser polarization also serves to reduce magnetic dipole (M1) contributions to the signal. While we are not aware of any calculations of the M1 amplitude for these transitions, it is expected to be small since the radial wavefunctions of the 2s, 3s, and 4s states are orthogonal to one another. If relativistic effects are large enough to allow a weak M1 amplitude, this effect can be reduced by use of counter-propagating waves in a standing wave optical cavity.  Use of standing wave field for this purpose, as well as for enhancement of the PNC amplitude, were used in Refs.~\cite{Wood1997,Antypas2018.APV-Yb}. Reduction of systematic effects for the general coherent control process are discussed in more detail in Ref.~\cite{antypas2014measurement}.
}

There are a couple of options available for detection of atoms that have transitioned to the $3s$ state.  Note that, in addition to its sensitivity to $3s$ atoms, the detection method must be highly insensitive to atoms that have been excited to the $3p_{1/2}$ state.  The primary spontaneous decay paths for $3s$ atoms is to the $2p_{1/2}$ and $2p_{3/2}$ levels, followed by rapid decay to the $1s$ ground state.  Detection of the Lyman alpha radiation of the $2p_j \rightarrow 1s$ line, therefore seems a promising detection method for this measurement.  Detection of the fluorescence at 656 nm from the first leg of this decay does not seem feasible, as it suffers from the likely-copious presence of scattered laser light, and also from the spontaneous fluorescence resulting from the decay of $3p_{1/2}$ atoms to the $2s$ state, both of which would be at similar wavelengths, and therefore difficult to separate.  The $3p_{1/2}$ state can decay directly to the $1s$ ground state, or to the $2s$ state.  The 102.6 nm fluorescence due to the former must, therefore, be blocked, but the latter is not expected to interfere with the Lyman alpha detection, as the $2s$ state is long lived (lifetime $\sim$ 100 ms). A second detection option would be to apply a weak probe beam tuned to the $3s \rightarrow 2p_{3/2}$ line to the interaction region, and observe the gain in this probe beam. We choose this line as it is distinct from the interfering transitions used to populate the $3s$ level, in contrast to the detection method proposed in Ref.~\cite{Rasor2020}. The frequency of this line differs from that of the excitation wavelength by $\sim$10 GHz.  This probe laser intensity should be maintained at a level much less than the saturation intensity of the transition so as to not affect the $3s$ state population significantly. We note that the frequency of the 656 nm excitation beam is sufficient to photoionize the $3s$ state through a one-photon interaction.  A detection scheme based on photoionization, however, is not feasible, as photoions generated from the $3p_{1/2}$ state would mask the $3s$ signal.  In fact, this photoionization rate should be minimized to the extent possible, to avoid space charge effects that might otherwise obscure the signal.  

Two pathway coherent control allows for the measurement of very weak transition moments such as $E_{\textrm{PNC}}$, referenced against the Stark transition polarizability. The limiting systematic effect encountered in the previous microwave approach resulting from the mechanical rotation of a microwave cavity would not contribute to this optical measurement.   
As shown in this section, this all-optical approach can provide a precise measurement of $C_{p}^{(1)}$ and $C_{p}^{(2)}$. Given that these constants can be extracted simultaneously, we remind the reader that the axial charge $g_A$, determined by the $C_p^{(2)}/ C_p^{(1)}$ ratio is an important quantity in understanding the spin structure of the nucleon, and it also plays a role in certain astrophysical processes. It can be measured in the $\beta$ decay of the neutron. Its  current recommended value~\cite{Workman:2022ynf} is $g_A = 1.2756(13)$. This value can provide a consistency check for the $C_p^{(2)}/ C_p^{(1)}$ ratio.

%

\section{Conclusion}
A long standing goal in atomic parity violation has been a measurement in hydrogen.  The attraction of hydrogen is two-fold.  First, the electronic structure of hydrogen is much better known than that of any other atom, greatly facilitating interpretation of the results.  Second, since the hydrogen nucleus contains only a single proton, measurement of the nuclear spin-independent coupling coefficient $C_{p}^{(1)}$ leads directly to ${\sin }^2{\theta_\mathrm{W}}$.  Prior measurements based on microwave transitions between hyperfine components of the $2s_{1/2}$ state were ultimately not successful.  In the present work, we have outlined how similar measurements carried out on the $2s_{1/2} \rightarrow 3s_{1/2}$ or $2s_{1/2} \rightarrow 4s_{1/2}$ transition might prove fruitful, and described an experimental technique for such an investigation.

\acknowledgments
We are happy to acknowledge a useful discussion with R. Dunford.  This work  was supported by the U.S. National Science Foundation under Grant Number PHY-1912519 and PHY-2207546, and by the Sara Louise Hartman endowed professorship in Physics. J.L. is grateful to the University of Nevada, Reno for its hospitality and acknowledges the financial support by the National Natural Science Foundation of China (Grant No. 11874090).

\bibliography{library-apd,PNC}

\end{document}